\documentclass[%
 reprint, 
nofootinbib,
 amsmath,amssymb,
 aps, physrev,
floatfix,
]{revtex4-2}

\usepackage{graphicx}
\usepackage{dcolumn}
\usepackage{bm}
\usepackage{xcolor}
\usepackage{verbatim}
\usepackage[normalem]{ulem} 

\usepackage[rightcaption]{sidecap}

\usepackage{url}

\DeclareMathAlphabet\mathbfcal{OMS}{cmsy}{b}{n}

\usepackage[hidelinks]{hyperref}
\begin{document}

\preprint{APS/123-QED}

\title{Measurement of the Dispersion--Galaxy Cross-Power Spectrum with the Second
CHIME/FRB Catalog
}


\author{Haochen~Wang$^{1,2}$}
\email[]{hcwang96@mit.edu}
\author{Kiyoshi~Masui$^{1,2}$}
\author{Shion~Andrew$^{1,2}$}
\author{Emmanuel~Fonseca$^{3,4}$}
\author{B. M.~Gaensler$^{5,6,7}$}
\author{R. C.~Joseph$^{8,9}$}
\author{Victoria M.~Kaspi$^{8,9}$}
\author{Bikash~Kharel$^{3,4}$}
\author{Adam E.~Lanman$^{1,2}$}
\author{Calvin~Leung$^{10, 11}$}
\author{Lluis~Mas-Ribas$^{5}$}
\author{Juan~Mena-Parra$^{6,7}$}
\author{Kenzie~Nimmo$^{2}$}
\author{Aaron B.~Pearlman$^{8,9}$}
\author{Ue-Li~Pen$^{12,13,14,6,15}$}
\author{J. Xavier~Prochaska$^{5, 16, 17}$}
\author{Ryan~Raikman$^{1,2}$}
\author{Kaitlyn~Shin$^{1,2}$}
\author{Seth R.~Siegel$^{15, 8, 9}$}
\author{Kendrick M.~Smith$^{15}$}
\author{Ingrid H.~Stairs$^{18}$}

\affiliation{$^1$ Department of Physics, Massachusetts Institute of Technology, 77 Massachusetts Avenue Cambridge, MA 02139, USA}
\affiliation{$^2$ MIT Kavli Institute for Astrophysics and Space Research, Massachusetts Institute of Technology, 77 Massachusetts Avenue Cambridge, MA 02139, USA}
\affiliation{$^3$ Department of Physics and Astronomy, West Virginia University, PO Box 6315, Morgantown, WV 26506, USA}
\affiliation{$^4$ Center for Gravitational Waves and Cosmology, West Virginia University, Chestnut Ridge Research Building, Morgantown, WV 26505, USA}
\affiliation{$^5$ Department of Astronomy and Astrophysics, University of California, Santa Cruz, 1156 High Street, Santa Cruz, CA 95064, USA}
\affiliation{$^6$ Dunlap Institute for Astronomy \& Astrophysics, University of Toronto, 50 St.~George Street, Toronto, ON M5S 3H4, Canada}
\affiliation{$^7$ David A.~Dunlap Department of Astronomy \& Astrophysics, University of Toronto, 50 St.~George Street, Toronto, ON M5S 3H4, Canada}
\affiliation{$^8$ Department of Physics, McGill University, 3600 rue University, Montr\'eal, QC H3A 2T8, Canada}
\affiliation{$^9$ Trottier Space Institute at McGill University, 3550 rue University, Montr\'eal, QC H3A 2A7, Canada}
\affiliation{$^{10}$ Department of Astronomy, University of California, Berkeley, CA 94720, United States}
\affiliation{$^{11}$ Miller Institute for Basic Research, University of California, Berkeley, CA 94720, USA}
\affiliation{$^{12}$ Institute of Astronomy and Astrophysics, Academia Sinica, Astronomy-Mathematics Building, No. 1, Sec. 4, Roosevelt Road, Taipei 10617, Taiwan}
\affiliation{$^{13}$ Canadian Institute for Theoretical Astrophysics, 60 St.~George Street, Toronto, ON M5S 3H8, Canada}
\affiliation{$^{14}$ Canadian Institute for Advanced Research, 180 Dundas St West, Toronto, ON M5G 1Z8, Canada}
\affiliation{$^{15}$ Perimeter Institute for Theoretical Physics, 31 Caroline Street N, Waterloo, ON N25 2YL, Canada}
\affiliation{$^{16}$ Kavli Institute for the Physics and Mathematics of the Universe (Kavli IPMU), 5-1-5 Kashiwanoha, Kashiwa, 277-8583, Japan}
\affiliation{$^{17}$ Division of Science, National Astronomical Observatory of Japan, 2-21-1 Osawa, Mitaka, Tokyo 181-8588, Japan}
\affiliation{$^{18}$ Department of Physics and Astronomy, University of British Columbia, 6224 Agricultural Road, Vancouver, BC V6T 1Z1 Canada}

\date{\today}

\begin{abstract}



The dispersion of extragalactic fast radio bursts (FRBs) can serve as a
powerful probe of the diffuse plasma between and surrounding galaxies,
which contains most of the Universe's baryons.
By cross-correlating the dispersion of background FRBs with the locations
of foreground galaxies, we can study the relative spatial distributions of plasma and
galaxies on scales of 0.1 to 50\,Mpc, which are strongly affected by
feedback processes in galaxy formation. Here we present the measurement of
the dispersion--galaxy angular cross-power spectrum between 2873 FRBs from the Second
CHIME/FRB Catalog and nearly 6 million galaxies from the Dark Energy Spectroscopic
Instrument (DESI) Legacy Imaging Survey. Over five photometric galaxy
redshift bins spanning $0.05 < z <0.5$ and at 5.1$\sigma$
significance, we make the first definitive detection of spatial correlations in
FRB dispersion measure due to cosmic structure.
While parameter inferences should be interpreted with
caution because of incomplete modelling of both the signal and systematic
errors, our data indicate that the plasma--galaxy cross-power spectrum cuts
off relative to the matter power spectrum at a scale 
$k_\textrm{cut}^{-1}=0.9^{+0.4}_{-0.4}\,\textrm{Mpc}$.
This scale is consistent with those X-ray stacking analyses that suggest dark-matter halos with group-scale masses are largely evacuated of their baryons by feedback processes.
Our study demonstrates that FRBs are promising tools to discern the physics of 
baryonic structure formation and will only become more powerful as FRB
surveys expand.


\end{abstract}

\maketitle



\emph{Introduction and theory.}---The dispersion of fast radio bursts (FRBs) provides a unique opportunity to study the spatial distribution of baryons in the Universe \citep{Lorimer_2007, review_2022}. As these millisecond-duration extragalactic flashes propagate through diffuse plasma, they acquire a 
frequency-dependent delay in arrival time that can be precisely measured upon
detection by a radio telescope. The extent of this pulse dispersion depends on distance and line-of-sight electron number density, and is quantified by the dispersion measure (DM):
\begin{equation} \label{eq:dm_def}
    \text{DM}(\hat n, z) \equiv \int_0^{\chi(z)} \,d\chi^\prime
    \,\frac{n_e(\hat n, \chi^\prime)}{[1+z(\chi^\prime)]^2},
\end{equation}
where $n_e$ is the free-electron number density and $\hat n$ and
$\chi$ are the direction and comoving distance to the FRB, respectively.
While the Milky Way, the FRB host, and the FRB environment all contribute to
the DM, a substantial ``cosmic''
component comes from the intergalactic medium (IGM) and gas surrounding intervening
galaxies \citep[e.g.][]{missing_baryons, Prochaska_2011, Nevalainen_2015}.
Since such gas accounts for about 90\% of the Universe's 
baryons---far out-weighing stars or the cooler gas
within galaxies---the cosmic dispersion is an
excellent tracer of the total baryonic matter \citep{McQuinn_2014}.

Intergalactic baryons are difficult to study by other means. The
main observational handles include the thermal Sunyaev--Zeldovich effect
\citep[e.g.][]{2019MNRAS.483..223T, 2019A&A...624A..48D}, kinetic Sunyaev--Zeldovich effect \citep[e.g.][]{guachalla2025, sunseri2025}, and free--free
emission of X-rays \citep[e.g.][]{pen1999, 2018Natur.558..406N}, all of which are primarily sensitive to the
densest plasma, especially galaxy clusters, leaving more diffuse structures
poorly constrained. Quasar absorption lines can be used to study baryons in
the IGM, but the high temperature and low neutral fraction at low redshift result in shallow and broad absorption features, making their detection and interpretation challenging \citep[e.g.][]{Tejos_2016, pessa2025}. In simulations, baryons are strongly affected by the small-scale
astrophysics of supernova and active galactic nuclei feedback in galaxy
formation, both of which are poorly understood \citep[e.g.][]{sorini_2022, Ayromlou_2023, Khrykin_sim}. The sensitivity of baryons to feedback processes, in turn, provides an
opportunity: should we measure the spatial distribution of baryons, we can
constrain feedback models and better understand galaxy assembly and evolution.

To date, studies of cosmic baryons using dispersion have mainly considered the
probability distribution function of DM as a function of FRB redshift, with
wider distributions indicative of a clumpier spatial distribution
\citep{McQuinn_2014, Macquart_2020,  Baptista_2024, Medlock_2024, medlock2025, sharma2025}. While highly sensitive to
baryon physics, this statistic lacks spatial information and thus requires
detailed modelling to connect it to the baryon distribution on different
spatial scales \citep{sharma2025}. Another approach is stacking analyses involving
DM and galaxies, which have yielded $\sim3\sigma$ detections of excess DM
associated with galaxies on 0.1- to $5$-Mpc scales
\citep{McQuinn_2014, connor2022, Wu_2023, flimflam, hussaini2025}. Other works have
cross-correlated FRB and galaxy positions or statistically
associated FRBs with host galaxy clusters \citep{Rafiei-Ravandi_2021,
Rafiei-Ravandi_2024}, but no constraints on their baryonic contents were
made.

In contrast,
angular correlation statistics involving FRB DMs depend on the free-electron power
spectra, encoding spatial information about the baryon distribution
\citep{masui2015, shirasaki2017, madhavacheril2019, alonso2021}.
This is the approach taken here, where we cross-correlate the
ionized gas, as traced by the FRB DM, with galaxy positions
to produce the dispersion--galaxy angular cross-power spectrum. This is
possible with the Second CHIME/FRB Catalog (Catalog 2 hereafter), the largest FRB catalog
observed by a single instrument to date, with the Dark Energy Spectroscopic
Instrument (DESI) Legacy Imaging Survey (LIS hereafter) providing the galaxy catalog.

Specifically, each FRB samples the DM field in Eq.~(\ref{eq:dm_def}) at
a discrete point in space.  We are principally interested in the cosmic
contribution, so
for a given FRB from Catalog 2 (with label $i$, dispersion DM$_i$, and sky
location $\hat n_i$), we subtract the \texttt{NE2001} model for the Milky Way contribution
\citep{ne2001} to obtain our input measurements:
\begin{align}
    d(\hat{n}_i) &= \text{DM}_i - \text{DM}_\text{MW}(\hat n_i),
\end{align}
which samples the 3-dimensional DM field defined by Eq.~(\ref{eq:dm_def}) stochastically over the redshift, and we define the DM overdensity field
\begin{align}
    \label{eq:dm_overd_def}
    \Delta d (\hat{n}_i) &\equiv d(\hat{n}_i) - \bar{d},
\end{align}
with $\bar{d}$ denoting the average of $d(\hat{n}_i)$ over all FRBs in the sample.

Similarly, after binning in redshift $z_g$, the galaxy catalog can be used to
construct the 2D galaxy number density $ n^{(2)}_g(\hat n, z_g)$, which
samples from the underlying overdensity:
\begin{equation}
    \Delta g (\hat{n}, z_g) = n^{(2)}_g(\hat n, z_g) / \bar n^{(2)}_g(z_g) - 1,
\end{equation}
where $\bar
n^{(2)}_g$ is the average galaxy density.



Since galaxies and the cosmic plasma should be clustered on large scales,
the DM and galaxy overdensities correlate as
\begin{equation}
    \langle \Delta d_{lm} \Delta g_{l^\prime m^\prime}^*(z_g) \rangle =
    \delta_{l \, l^\prime} \delta_{m \, m^\prime} C_l^{dg}(z_g), 
\end{equation}
where $\Delta d_{lm}$ and $\Delta g_{lm}$ are the spherical
harmonic transforms of their respective 2D fields, $l$ and $m$ label the
spherical harmonics, $\delta_{l \, l^\prime}$ and $\delta_{m \, m^\prime}$ are Kronecker delta functions, and $C_l^{dg}$ is the dispersion--galaxy angular cross-power spectrum.
In the Supplementary Material, we use the flat-sky and Limber approximations to
derive a model for $C_l^{dg}$, showing that
\begin{widetext}
\begin{equation} \label{eq:cl_dg}
    C_l^{dg}(z_g) = \frac{n_{e}(z_g)}{(1 + z_g)^2} \frac{f_f(z_g)}{\chi_g^2} P_{eg}(k = l/\chi_g, z_g)
    + \left[ \langle \text{DM}_{\text{M}} \rangle  + \langle \text{DM}_{\text{C}} (z_g) \rangle +\frac{\langle \text{DM}_\text{H} \rangle}{1 + z_g} - \bar{d} \right]
    \frac{p_f(z_g)}{\chi_g^2} H(z_g) P_{fg}(k = l/\chi_g, z_g),
\end{equation}
\end{widetext}
where $n_{e}(z_g)$ is the electron number density at $z=z_g$, $\chi_g$
is the comoving distance to the galaxy redshift $z_g$, $p_f(z_g)$ is the probability of detecting an FRB at redshift $z_g$, $f_f(z_g) \equiv
\int_{z_g}^\infty dz\,p_f(z)$ is the
fraction of FRBs located in the background of the galaxies, $P_{eg}(k,z)$
is the electron-galaxy cross-power spectrum, $\langle
\text{DM}_{\text{M}} \rangle$ is the mean Milky Way halo DM contribution, $\langle
\text{DM}_{\text{C}} (z) \rangle$ is the mean DM as a function of
redshift (known as the Macquart relation \cite{Macquart_2020}), $\langle
\text{DM}_\text{H} \rangle$ is the mean FRB host DM contribution, $H(z)$ is the
Hubble expansion rate, and $P_{fg}(k,z)$ is the cross-power spectrum between
FRB and galaxy number counts. Note that the mean Milky Way halo DM contribution $\langle
\text{DM}_{\text{M}} \rangle$ is poorly measured. We adopt $\langle
\text{DM}_{\text{M}} \rangle = 80 \text{ pc} \text{ cm}^{-3}$ to be consistent with studies on the Galactic halo \citep{halo_2019, halo_2020}.

The first term in the above equation results from the clustering of electrons
with galaxies when both are foreground to the FRBs. The first term is positive,
since electrons and galaxies are expected to be correlated. The
second term comes from the galaxies clustering with the FRBs themselves when
at the same redshift. Note from its first factor, this term may be
positive or negative, depending on whether the FRBs at that redshift typically
have more or less than the average dispersion.
Fitting Eq.~(\ref{eq:cl_dg}) to a measurement of
the dispersion--galaxy cross-power spectrum from data can constrain galaxy feedback through the electron-galaxy
cross-power spectrum $P_{eg}(k,z)$.



\begin{figure}
\includegraphics[width=1\columnwidth]{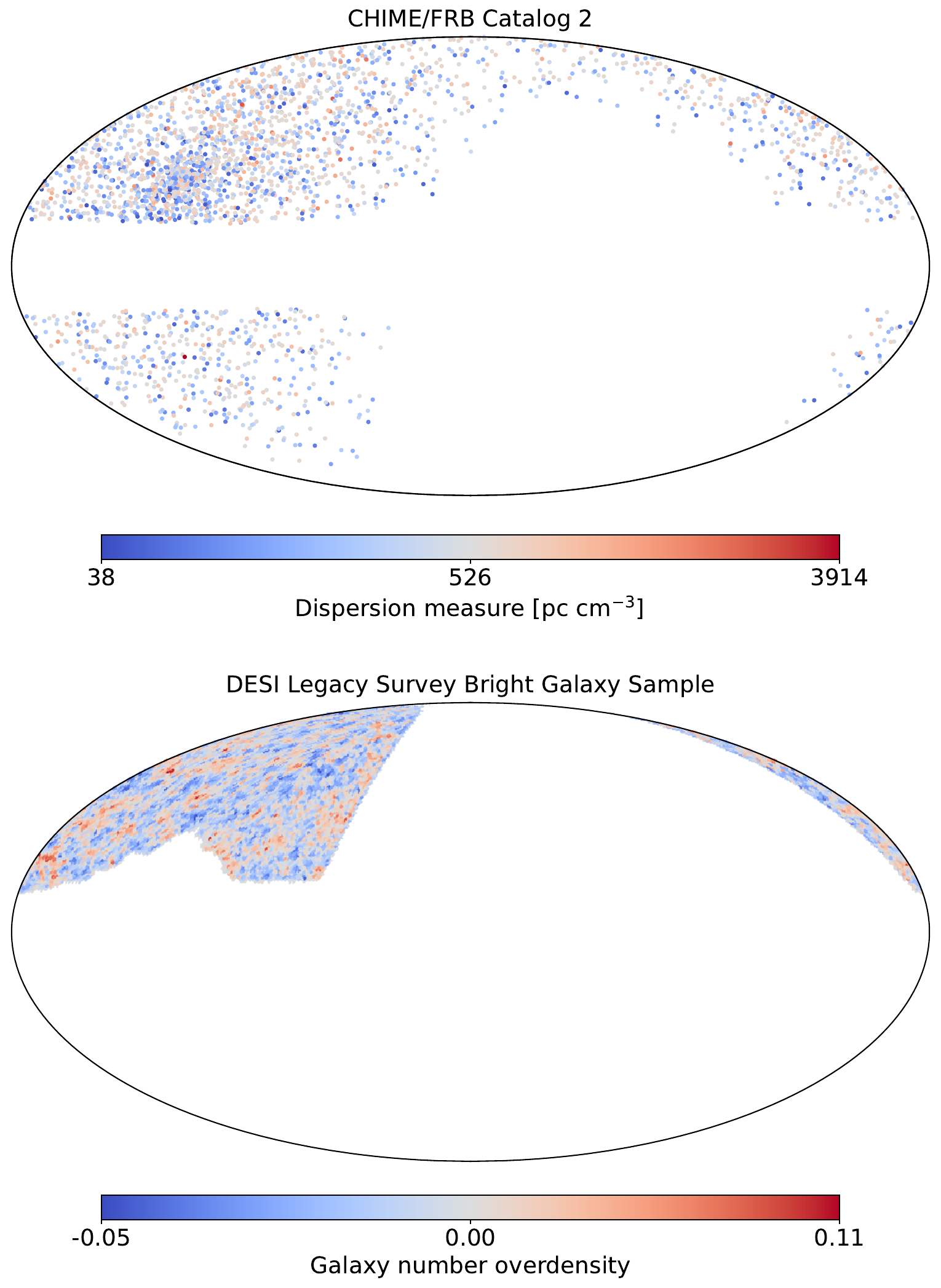}
\caption{\label{fig:catalogs}Catalogs used in this work. Top: DM from the CHIME/FRB Catalog 2. FRBs detected only in side lobes and FRBs with high Galactic DM contributions ($\text{DM}_\text{MW} > 100 \text{ pc} \text{ cm}^{-3} $ predicted by \texttt{NE2001}) have been excluded. Bottom: DESI LIS BGS galaxy number overdensity in the North field. Only the $0.1 < z_g < 0.2$ photometric redshift slice is shown. Both panels are shown in the galactic coordinates.}
\end{figure}

\emph{Methods.}---
Observing the 400-800 MHz radio band, the
Canadian Hydrogen Intensity Mapping Experiment (CHIME) is a radio telescope
that scans through the north hemisphere every day \cite{chime_overview}. The Second CHIME/FRB
Catalog, containing 4547 FRBs typically localized with $0.2^\circ \times
0.2^\circ$ angular resolution, is currently the largest FRB sample collected by a single
instrument \cite{chime_cat_2}. To mitigate the large uncertainties of the Galactic DM
contribution near the Galactic plane, we remove FRBs with
$\text{DM}_{\text{MW}} > 100 \text{ pc } \text{cm}^{-3}$ as predicted by the
\texttt{NE2001} model. All DM values used in this study are measured with \texttt{fitburst} \citep{Fonseca_2024} and have their \texttt{NE2001} Galactic DM contributions subtracted. We also exclude FRBs that were detected only through
side lobes \citep{chime_cat_2}, due to their poor localizations. For FRBs with multiple subbursts
and repeaters, we only use the position and DM value from their first
burst/occurrence for simplicity (note that 
the variation in DM
between detections is at the percent level). There are 2873 unique
FRB sources in our sample after these cuts (Fig.~\ref{fig:catalogs}, top panel). 

The Dark Energy Spectroscopic Instrument (DESI) \cite{desi_dr1} is currently
collecting data and is the largest spectroscopic redshift survey to date. The DESI Legacy Imaging
Survey (LIS) \cite{Dey_2019} is the photometric survey that provides targets
for DESI spectroscopic follow-ups. To identify low-redshift foreground galaxies to
cross-correlate with FRBs, we select the Bright Galaxy Samples (BGS) from LIS using
DESI's color selection criteria with the corresponding data quality masks (see details in the
Supplementary Material). We use BGS in the North field of LIS Data Release 8 for this analysis,
which include nearly 6 million galaxies between photometric redshift 0.05 and 0.5
(Fig.~\ref{fig:catalogs}, bottom panel). The North field overlaps roughly with
twice as many FRBs as the South field, which is observed with a different
instrument. To simplify data processing and control systematics, we use only
the North field for this analysis. To account for the survey geometry, we assemble a random catalog for the North BGS field using random survey objects generated by DESI\@. We further group the BGS galaxies into
five photometric redshift bins with edges 0.05, 0.1, 0.2, 0.3, 0.4, and 0.5,
noting that the typical redshift error is $\sigma_z \sim 0.03$. We will cross-correlate all of our FRBs with each galaxy redshift bin to extract redshift-sensitive information.

The DESI Collaboration has recently released the first year data of their
spectroscopic survey (DESI DR1) \citep{desi_dr1}, which contains over 4 million BGS galaxies across the entire DESI survey. 
Despite having accurate spectroscopic redshift measurement, DESI DR1 suffers from highly nonuniform survey geometry and incomplete sky coverage. We therefore only use DESI DR1 data to perform consistency checks (see \emph{Fits and validations}).

To measure the dispersion--galaxy angular cross-power spectrum from the
two
catalogs, we use the \texttt{NaMaster} Python package developed by
\cite{Alonso_2019}. More specifically, we use its catalog-based estimator
\cite{Wolz_2025}, which has the advantage of not needing to pixelize the catalog
fields using a pixelization scheme such as HEALPix. Following the procedure
laid out in \cite{tutorial}, we create an \texttt{NmtFieldCatalog} object to
store DM catalog information by providing the angular positions of FRBs and
their corresponding DM overdensity values (defined in
Eq.~[\ref{eq:dm_overd_def}]) with a uniform weight. To compute and store the
galaxy overdensities, we create an \texttt{NmtFieldCatalogClustering} object by
providing the angular positions of galaxies from the real and random catalogs.
For DESI LIS, no weights are provided for the galaxies, so we adopt a uniform
weight for both the real and random catalog. For DESI DR1 data, we use the
weights provided by DESI\@. We use \texttt{compute\_coupled\_cell} to compute the angular cross-power spectrum from the two catalog objects and then deconvolve the effect of the survey window functions on the power spectrum using the \texttt{NmtWorkspace} object with a binning scheme specified in the section below.

\begin{figure*}[htb]
\includegraphics[width=2\columnwidth]{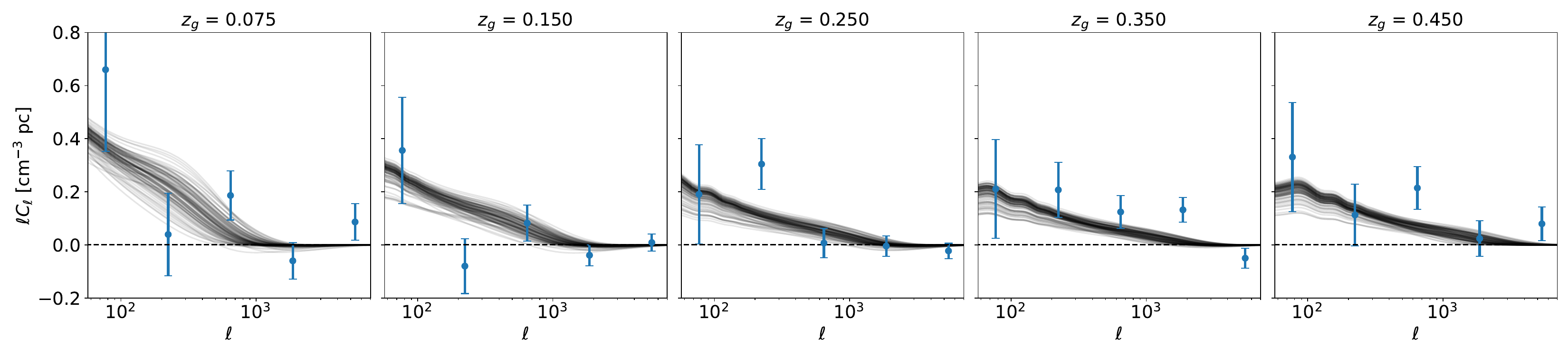}
\caption{\label{fig:simple_fit} The dispersion--galaxy cross-power spectra (blue data points) measured with 5 equally spaced log bins over the range $40 < l < 8000$, between the same FRB samples and galaxy bins centered at five different photometric redshifts. The theoretical model (Eq.~[\ref{eq:cl_dg}]) is fit simultaneously to the power spectra with 10 logarithmic $l$ bins to obtain the parameter posterior distribution. We also show curves computed from the theoretical model (black), using 100 randomly selected sets of parameter values from the posterior distribution to illustrate the allowed shapes and amplitudes of the fit signal.}
\end{figure*}

\emph{Results.}---Figure~\ref{fig:simple_fit} shows the measured dispersion--galaxy angular cross-power spectra. Each measurement is made between the DM of the same 2873 FRBs from Catalog 2 and a different galaxy photometric redshift bin from the Legacy Survey. In Fig.~\ref{fig:simple_fit}, we present each power spectrum using five equally spaced logarithmic bins over the range $40 < \ell < 8000$ to ensure a clean visualization. However, model parameters are constrained using power spectra measured with ten logarithmic bins over the same $\ell$ range to minimize the effect of the power spectrum bandpower window (see \emph{Fits and validations}). Note that we have excluded the measurements for $\ell < 40$ since these measurements are susceptible to large-scale bias from photometric depth variations in the galaxy survey and the Galactic DM variations in the FRB catalog, which we will examine in \emph{Fits and validations}. 

To fit the theory template Eq.~(\ref{eq:cl_dg}) to the measurement, we evaluate
Eq.~(\ref{eq:cl_dg}) at the central redshift $z_g$ of each photometric redshift
bin and compute its corresponding comoving distance $\chi_g$ using the
Planck-measured cosmology \cite{planck}. However, note that the first two
redshift bins span their redshift ranges by a factor of two (from $z_g = 0.05$
to $z_g = 0.1$ and from $z_g = 0.1$ to $z_g = 0.2$). To increase the accuracy
of the model for these two bins, we subdivide each bin into two subbins equally
spaced in redshift and sum the templates evaluated at the central redshift of
each subbin, weighted by the number of galaxies in each subbin.

We adopt $n_{e0} = 1.86\times10^{-7} \text{cm}^{-3}$ from the \texttt{FRB} software package \citep{FRB_software} as the comoving electron number density, and obtain the physical electron number density $n_e(z_g) = n_{e0} (1 +z_g)^3$. The mean IGM DM contribution, $\langle \text{DM}_{\text{C}} (z_g) \rangle$, is evaluated using the Macquart relation \cite{Macquart_2020}, and $\bar{d} = 623 \text{ pc } \text{cm}^{-3}$ is the average DM value of all FRBs. We model the FRB--galaxy cross-power spectrum using a simple linear bias model \citep{BBKS} as
\begin{align}
    & P_{fg}(k, \chi_g) = b_f b_g(z_g) P_m(k, \chi_g),
\end{align}
where $b_f$ and $b_g(z_g)$ are the linear bias factors for FRB numbers and
galaxies, respectively, and $P_m(k, \chi_g)$ is the Halofit non-linear matter power spectrum
computed using \texttt{CAMB} and the Planck cosmology \citep{camb, Mead_2020}. We take the BGS bias values from \cite{desi_design}, which were modeled by the DESI team using data-informed simulations, and fit the value of $b_f$ to the data. We similarly model the electron--galaxy cross-power spectrum as
\begin{align}
    P_{eg}(k, \chi_g) = b_e b_g(z_g) P_m(k, \chi_g) e^{-k/k_{\text{cut}}},
    \label{e:peg}
\end{align}
where $b_e$ is the electron bias and is assumed to be 1 \citep{masui2015}, and
$e^{-k/k_{\text{cut}}}$ is a simple model for the cutoff on the
electron--galaxy power spectrum relative to the matter power spectrum with a characteristic scale $k_{\text{cut}}$
due to galaxy feedback.

To model the FRB redshift distribution $p_f(z_g)$ and the fraction of FRBs beyond the galaxy redshift $f_f(z_g)$, we assume that the number density of observed FRBs can be computed by integrating the Schechter luminosity function \citep{schechter} above a fixed threshold flux (see details in Supplementary Material), with two free parameters: the Schechter function power law index $\alpha$ and the horizon redshift $z_*$ corresponding to the maximum redshift from which we can detect an FRB with a typical luminosity. Finally, FRB localization errors will suppress the correlation strength between
the DM and the foreground galaxies, so we model this effect using a Gaussian beam cutoff as $(C_l^{dg})_{\text{model}} =
(C_l^{dg})_{\text{theory}} \exp(-l^2/2l_\textrm{loc}^2)$, where $(C_l^{dg})_{\text{theory}}$ is given by Eq.~(\ref{eq:cl_dg}) with the modeling choices described above.

\begin{figure*}
\includegraphics[width=1.9\columnwidth]{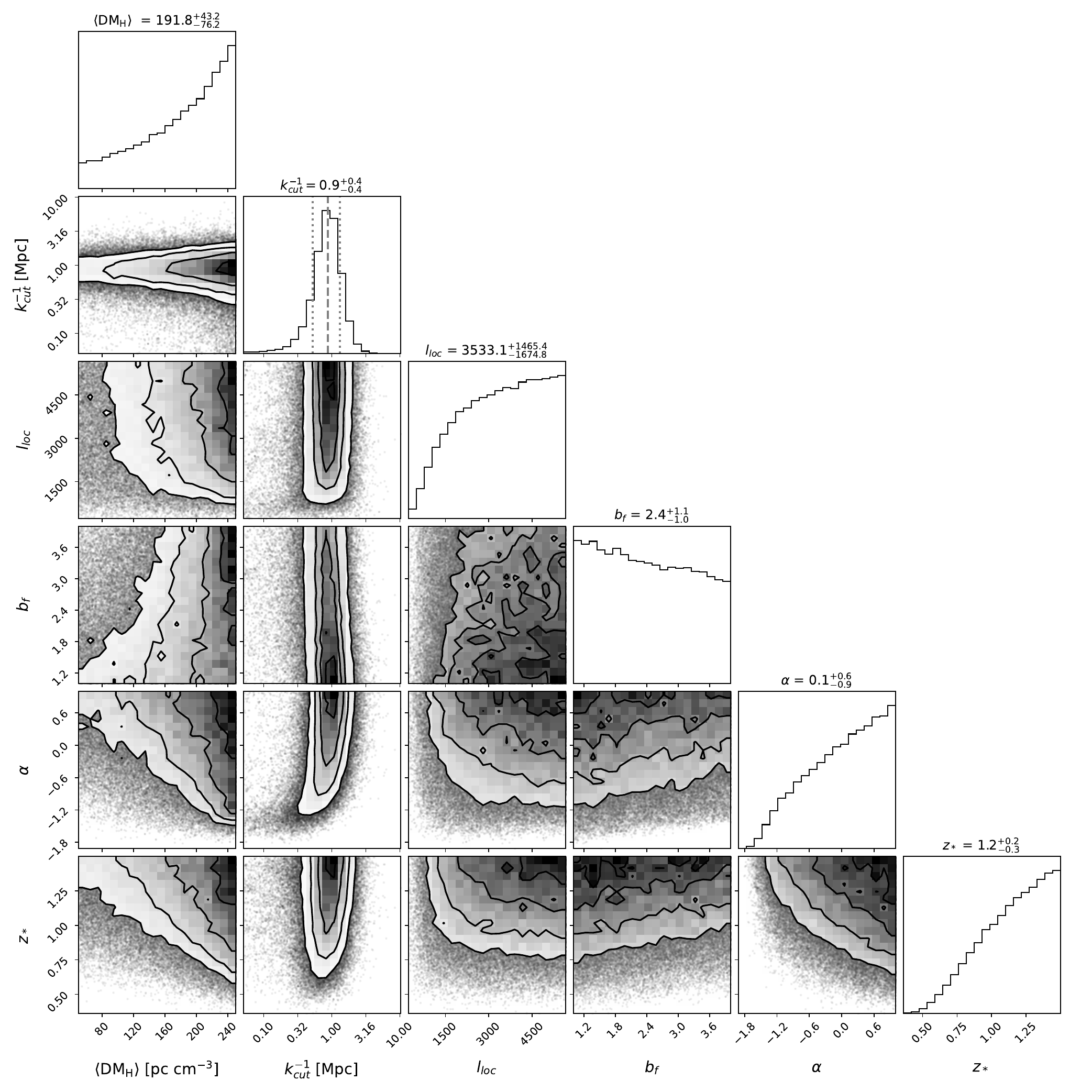}
\caption{\label{fig:posterior} Posterior distributions of the fit parameters. The 0.5-, 1-, 1.5-, and 2-sigma confidence regions are shown in the 2D distributions, and the 16th, 50th, and 84th percentiles are listed above each 1D marginalized distribution. The parameter $k^{-1}_{cut}$ parametrizes galaxy feedback strength and is well constrained. The power spectrum cutoff scale $l_{\text{loc}}$ is preferred over the range $l_{\text{loc}} > 1000$. All the other parameters are related to the FRB population and host environment, which our signal model is insensitive to.}
\end{figure*}

We use the \texttt{emcee} package to fit the theory template to the measured
dispersion--galaxy cross-power spectra over 10 logarithmic $l$ bins. For the comoving cutoff scale of the electron--galaxy power spectrum $k^{-1}_{cut}$, we adopt a logarithmic prior of $0.05 \text{ Mpc} < k^{-1}_{cut} < 50 \text{ Mpc}$ to sample its value across three orders of magnitude. For the rest of the parameters, we adopt a linear prior consistent with previous CHIME FRB population analyses \citep[e.g.][]{Rafiei-Ravandi_2021, Shin_2023}. The parameter priors and
posterior distributions of the fitting parameters are shown in \emph{Fits and validations} (
Fig.~\ref{fig:posterior}). Theoretical curves from 100 randomly
selected sets of parameters from the posterior are shown in
Fig.~\ref{fig:simple_fit}. 

With details in \emph{Fits and validations}, we find that the measured cross-power spectra are dominated by the electron--galaxy component (the $P_{eg}$ term of Eq.~[\ref{eq:cl_dg}]) and therefore insensitive to the FRB properties mostly affecting the $P_{fg}$ term. As a result, the feedback parameter $k^{-1}_{cut}$ is well constrained but not the parameters that model the FRB population. The large prior space volume that is consistent with a subdominant $P_{fg}$ contribution causes the skewness in the distributions of
$\langle \text{DM}_{\text{H}} \rangle$, $\alpha$, and $z_*$, but we observe that the median values of all parameters match their expectations from the literature \citep[e.g.][]{Shin_2023}. Lastly, we find that the cutoff scale on the power spectra due to FRB localization error $l_{\text{loc}}$ is preferred in the range over $l_{\text{loc}} > 1000$, consistent with CHIME's $\sim 0.2^\circ$ angular resolution and the constraints from \citep{Rafiei-Ravandi_2021}.

To determine the detection significance, we use the $\Delta \chi^2$ test to
reject an alternative null hypothesis assuming that the true signal is zero and that the measured power spectra are completely due to noise. We define $\Delta \chi^2 \equiv \chi_0^2 -
\chi_{min}^2$, where $\chi_0^2$ is computed between the null model and data,
and $\chi_{min}^2$ is the minimum $\chi^2$ between our model and data. This
quantity asymptotically follows the $\chi^2$ distribution with degrees of
freedom equal to the effective number of model parameters. Due to notable
degeneracies, the effective number of model parameters is lower than the actual
number of parameters in our model. Following \cite{bayesian_dim}, we use the
Bayesian model dimensionality estimated using $d_M = 2[\langle (\ln
\mathcal{L}^2) \rangle - \langle \ln \mathcal{L} \rangle^2]$, where
$\mathcal{L}$ is the likelihood function and the expectation $\langle \ldots \rangle$ is taken over the posterior, as the effective number of model parameters. We find that $\Delta \chi^2 = 31.2$ and $d_M = 2.4 \sim 2$. This results in a detection significance of 5.1$\sigma$. Using an alternative Bayesian test, we find a Bayes factor of $7\times10^4$, exceeding the commonly accepted threshold (Bayes factor $> 100$) for a decisive detection.


\emph{Discussion and conclusion.}---At $5.1\sigma$ significance, our results mark the first definitive
detection of spatial correlations in FRB dispersion from cosmic structure.
Previous works have seen analogous signals in stacking analyses (where FRB DMs are co-added as a function of impact parameters from
foreground galaxies), but at mixed statistical significance and smaller spatial scales \citep{connor2022,
Wu_2023, hussaini2025}. Other works have probed the cosmic baryons by fitting a parametric matter density model from a spectroscopic survey of galaxies along FRB sight lines \citep{Lee_2022, flimflam}, but this technique has only been applied to a few FRBs so far.
The cosmic structure of baryons has also been
studied through the DM distribution as a function of redshift, i.e., the
one-point function \citep{james2022, baptista2024, connor2024}. However,
the lack of spatial information in one-point
statistics means that significant modelling is required to connect such
measurements to cosmic structure \citep{sharma2025}, and risks contamination from
plasma local to the FRB\@. In contrast, our cross-correlation measurement isolates
the dispersion due to cosmic structure from the source-local contribution, and
provides a direct measurement of the power spectrum.

The model fits to our measured cross-power spectrum should be interpreted with
some caution due to the potential for systematic errors. While we have used
relatively wide redshift bins for the galaxies, we have not
modeled their photometric redshift errors. On the FRB side, selection effects
could alter the signal, especially the dispersion-dependent selection function
\citep{Qiu2025}
which has been measured to vary by a factor of 2 over the dispersions in our
sample \citep{cat_1}. Finally, our signal model in
Eq.~(\ref{eq:cl_dg}) could be incomplete, since its derivation requires
neglecting higher-order terms which may be significant on the scales included
in our measurement. 

\begin{figure} 
\includegraphics[width=1\columnwidth]{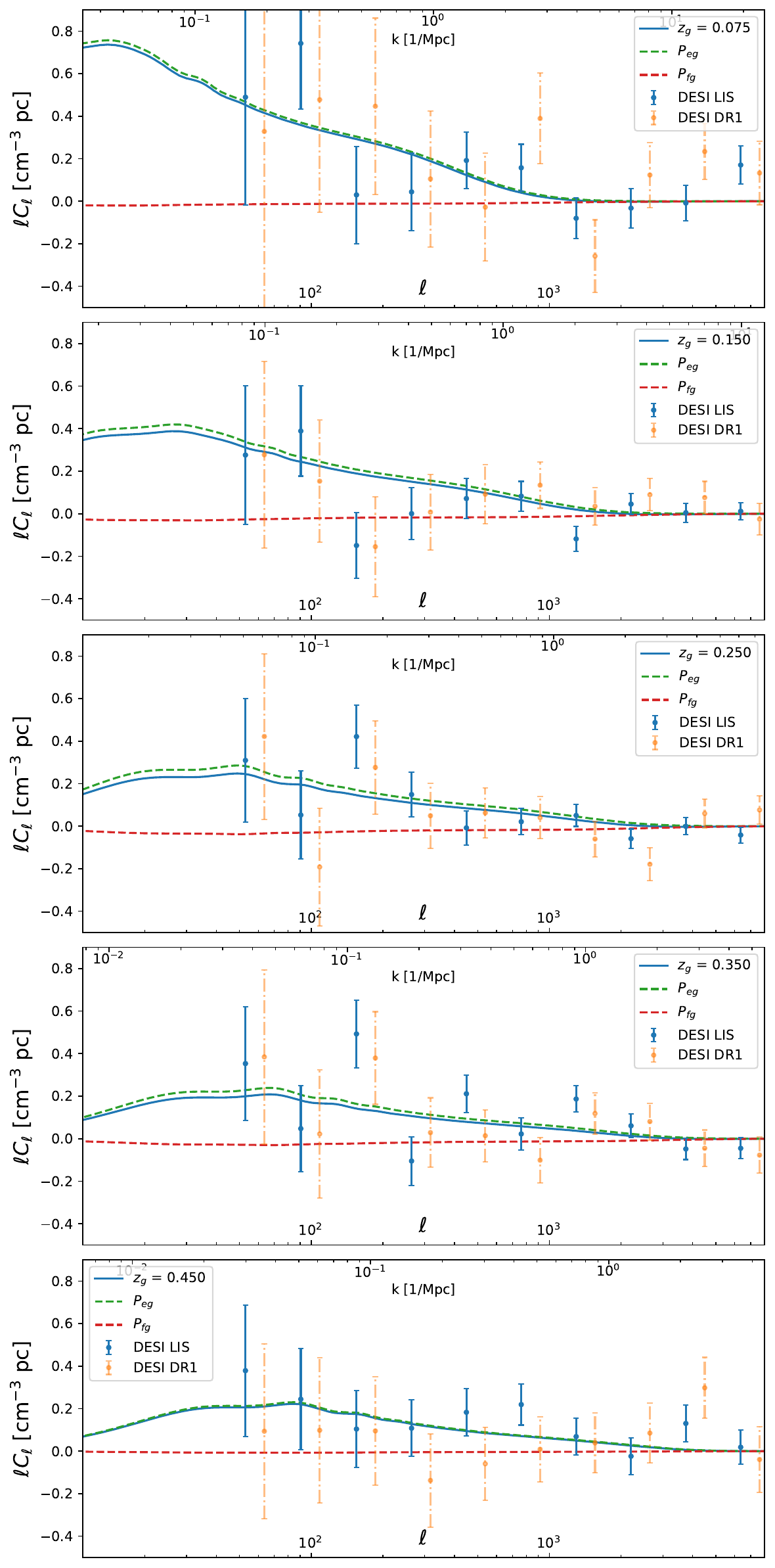}
\caption{\label{fig:full_fit} The dispersion--galaxy cross-power spectrum measured with 10 equally spaced log bins over $40 < l < 8000$ (blue data points), using galaxies centered at 5 different redshift bins. The signal template (solid curve) is computed from the mean parameter values of their posterior distributions (quoted at the top of Fig.~\ref{fig:posterior}), with the $P_{eg}$ and $P_{fg}$ components (dashed curves) plotted separately. As a consistency check, we also show the dispersion--galaxy cross-power spectrum measured using the DESI DR1 BGS sample (yellow data points).} 
\end{figure}

Our fits show that our model is consistent with the data for
parameter values well within their expected ranges. The one parameter that is well constrained is $k^{-1}_\textrm{cut}$: the scale at
which the electron--galaxy cross-power spectrum cuts off relative to the matter power spectrum. Here the power of our tomographic observations becomes apparent, since the single physical scale maps to a different angular scales at each redshift, as can be seen in Fig.~\ref{fig:full_fit}: $k \sim 1 \text{ Mpc}^{-1}$ where the attenuation on the power spectra begins due to feedback moves to higher $\ell$ at higher redshifts. This distinct feature in the power spectra means our model has very little freedom to
match the data; that they are nonetheless in agreement provides
confidence in both the robustness of our measurement and the validity of our
model. We also expect the systematics mentioned before more
strongly affect the power spectrum amplitudes than scales, so the scale-dependent power spectrum cutoff due to feedback, distinctively measured at various redshifts, makes the inference of $k^{-1}_\textrm{cut}$ more resilient to systematics. Additionally, $l_\textrm{loc}$ only weakly affects our model, since, as it
turns out, the power-spectra intrinsically cut off (roughly at $\ell \sim 300, 600,$ and 1000 for the first three redshift bins, respectively) on scales larger than
the localization-uncertainty limit ($l_{\text{loc}} > 1000$).

As such, our data indicate that, on
scales smaller than about 1\,Mpc, plasma does not strongly cluster with
galaxies, at least not those within DESI BGS\@. These galaxies have a
minimum halo mass of $\sim 10^{12}\,M_\odot$ and a mean halo
mass of $\sim2 \times 10^{13}\,M_\odot$ \citep{desi_design}---putting them in the mass
range of galaxy groups---and thus a viral radius of 0.6\,Mpc. In halos, dark matter
is concentrated well within the viral radius. However, some X-ray stacking analyses
have found, that even these relatively massive halos have expelled most of their baryons
via feedback processes \citep{2024A&A...690A.267Z, popesso2024}. Similar evidence has come from thermal Sunyaev--Zeldovich studies \citep{2019MNRAS.483..223T, 2019A&A...624A..48D}, where gas filaments are found to extend beyond galaxy pairs, and recent kinetic Sunyaev--Zeldovich works that have found evidence for enhanced feedback in the observed baryon density profiles \citep{guachalla2025,sunseri2025}.
Our measurements would seem to be in qualitative agreement with these works; however, much more
astrophysical modeling is required to interpret our measurement.

\begin{figure}[h]
\includegraphics[width=1\columnwidth]{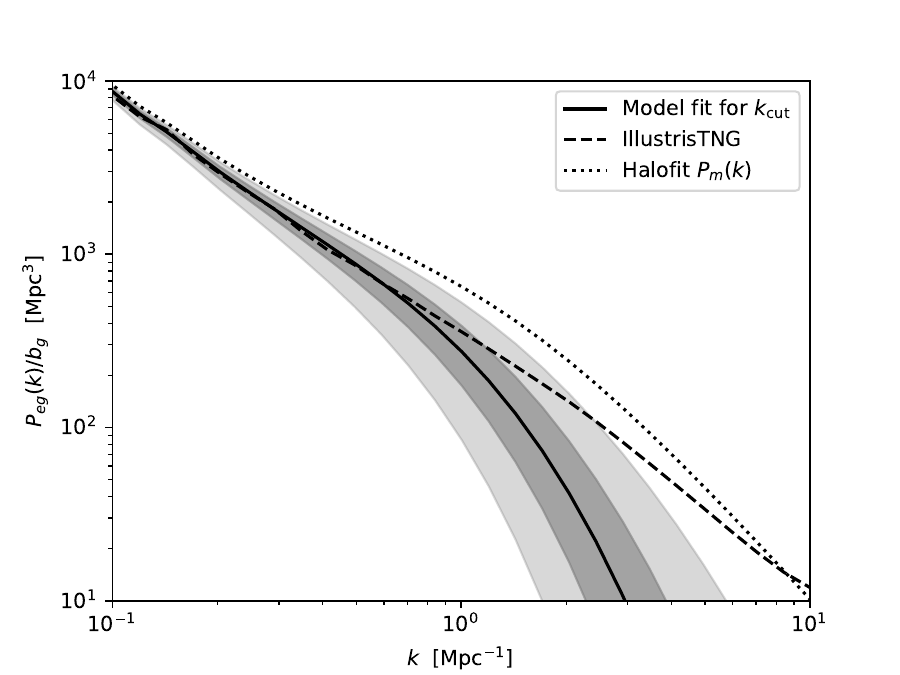}
    \caption{\label{fig:tngpk} Comparison of our fitted model for $P_{eg}(k)$
    to the prediction from the IllustrisTNG~300-1 simulation. Our model, given
    in Eq.~\ref{e:peg}, is the
    Halofit non-linear matter power spectrum (as calculated with \texttt{CAMB})
    with an exponential cutoff on small scales. The cutoff scale $k_\text{cut}$
    is the only free parameter and we plot the model for its median value and
    bands representing the 65\% and 95\% confidence intervals. For
    IllustrisTNG, we we show the cross-power spectrum between free electrons
    and ``subhalos'' with masses greater than $10^{12} M_\odot$. The Halofit
    matter power spectrum is also plotted for reference. All power
    spectra are calculated at $z=0.15$. Relevant factors
    of the galaxy bias are removed, and the electron bias is assumed to be
    unity. On scales larger than $k \sim 3\,\text{Mpc}^{-1}$, which 
    contain the bulk of our statistical
    power, $P_{eg}(k)$ is suppressed by
    an amount that agrees between our fitted model and IllustrisTNG.}
\end{figure}

To put our results in the context of cosmological structure formation
simulations, in Figure~\ref{fig:tngpk} we compare our model for 
$P_{eg}(k)$---with the inferred values for $k^{-1}_\textrm{cut}$---to the prediction from
the IllustrisTNG simulations \citep{2018MNRAS.475..624N, 2018MNRAS.477.1206N,
2018MNRAS.475..648P, 2018MNRAS.480.5113M, 2018MNRAS.475..676S}. At face value,
our fitted model agrees with IllustrisTNG up to $k \sim 3\,\text{Mpc}^{-1}$,
then diverges at smaller scales. While we have at least some statistical power out to
$k \sim 10\,\text{Mpc}^{-1}$, our model has an ad-hoc functional form and only one
parameter; thus a range of scales contribute to its constraints.
This, combined with the caveats about systematics, makes it unclear whether our
measurement is consistent with the simulation. We defer more thorough work
on modelling and inference to the future.

Measurements such as those presented here are poised to become more powerful
as FRB datasets mature. The number of detected FRBs will increase by
orders of magnitude with upcoming surveys by CHORD \citep{chord}, DSA-2000
\citep{DSA-2000}, and the SKA \citep{SKA_FRB}. However, even more significant will be the advent of large FRB
samples with optical redshifts \citep{chime_outrigger, kko_host,
dsa110_localized, ASKAP_localized}. Currently, the variance in our measured DMs
comes from both large-scale structure and
uncertainties in the FRB distances. Knowing FRB redshifts will allow us to
eliminate the latter. Redshifts would also allow us to only cross-correlate background
FRBs with foreground galaxies, eliminating the $P_{fg}$ term in
Eq.~(\ref{eq:cl_dg}) and isolating the $P_{eg}$ term to constrain feedback physics.
With these improvements to FRB data,
we expect cross-correlating FRB dispersion with other tracers of large-scale
structure to become the premier
method for studying Mpc-scale baryonic structure formation as shaped by galaxy-formation
feedback processes.



\emph{Acknowledgements.}---We thank Simon Foreman for insightful feedback and Liam Connor for helpful discussions on this study. We acknowledge that CHIME is located on the traditional, ancestral, and unceded territory of the Syilx/Okanagan people. We are grateful to the staff of the Dominion Radio Astrophysical Observatory, which is operated by the National Research Council of Canada. CHIME operations are funded by a grant from the NSERC Alliance Program and by support from McGill University, University of British Columbia, and University of Toronto. CHIME was funded by a grant from the Canada Foundation for Innovation (CFI) 2012 Leading Edge Fund (Project 31170) and by contributions from the provinces of British Columbia, Québec and Ontario. The CHIME/FRB Project was funded by a grant from the CFI 2015 Innovation Fund (Project 33213) and by contributions from the provinces of British Columbia and Québec, and by the Dunlap Institute for Astronomy and Astrophysics at the University of Toronto. Additional support was provided by the Canadian Institute for Advanced Research (CIFAR), the Trottier Space Institute at McGill University, and the University of British Columbia. 
E.F. is supported by the National Science Foundation (NSF) under grant number AST-2407399. K.W.M. holds the Adam J. Burgasser Chair in Astrophysics and recieved support from NSF grant 2018490. A.B.P. is a Banting Fellow, a McGill Space Institute~(MSI) Fellow, and a Fonds de Recherche du Quebec -- Nature et Technologies~(FRQNT) postdoctoral fellow. U.P.  is supported by the Natural Sciences and Engineering Research Council of Canada (NSERC), [funding reference number RGPIN-2019-06770, ALLRP 586559-23], Canadian Institute for Advanced Research (CIFAR), AMD AI Quantum Astro. K.S. is supported by the NSF Graduate Research Fellowship Program. J.M.P. acknowledges the support of an NSERC Discovery Grant (RGPIN-2023-05373). K.N. is an MIT Kavli Fellow. V.M.K. holds the Lorne Trottier Chair in Astrophysics \& Cosmology, a Distinguished James McGill Professorship, and receives support from an NSERC Discovery grant (RGPIN 228738-13). FRB research at UBC is supported by a NSERC Discovery Grant and by the Canadian Institute for Advanced Research.  This research used data obtained with the Dark Energy Spectroscopic Instrument (DESI). DESI construction and operations is managed by the Lawrence Berkeley National Laboratory. This material is based upon work supported by the U.S. Department of Energy, Office of Science, Office of High-Energy Physics, under Contract No. DE–AC02–05CH11231, and by the National Energy Research Scientific Computing Center, a DOE Office of Science User Facility under the same contract. Additional support for DESI was provided by the U.S. National Science Foundation (NSF), Division of Astronomical Sciences under Contract No. AST-0950945 to the NSF’s National Optical-Infrared Astronomy Research Laboratory; the Science and Technology Facilities Council of the United Kingdom; the Gordon and Betty Moore Foundation; the Heising-Simons Foundation; the French Alternative Energies and Atomic Energy Commission (CEA); the National Council of Humanities, Science and Technology of Mexico (CONAHCYT); the Ministry of Science and Innovation of Spain (MICINN), and by the DESI Member Institutions: www.desi.lbl.gov/collaborating-institutions. The DESI collaboration is honored to be permitted to conduct scientific research on I’oligam Du’ag (Kitt Peak), a mountain with particular significance to the Tohono O’odham Nation. Any opinions, findings, and conclusions or recommendations expressed in this material are those of the author(s) and do not necessarily reflect the views of the U.S. National Science Foundation, the U.S. Department of Energy, or any of the listed funding agencies.\ The DESI Legacy Imaging Surveys consist of three individual and complementary projects: the Dark Energy Camera Legacy Survey (DECaLS), the Beijing-Arizona Sky Survey (BASS), and the Mayall z-band Legacy Survey (MzLS). DECaLS, BASS and MzLS together include data obtained, respectively, at the Blanco telescope, Cerro Tololo Inter-American Observatory, NSF’s NOIRLab; the Bok telescope, Steward Observatory, University of Arizona; and the Mayall telescope, Kitt Peak National Observatory, NOIRLab. NOIRLab is operated by the Association of Universities for Research in Astronomy (AURA) under a cooperative agreement with the National Science Foundation. Pipeline processing and analyses of the data were supported by NOIRLab and the Lawrence Berkeley National Laboratory (LBNL). Legacy Surveys also uses data products from the Near-Earth Object Wide-field Infrared Survey Explorer (NEOWISE), a project of the Jet Propulsion Laboratory/California Institute of Technology, funded by the National Aeronautics and Space Administration. Legacy Surveys was supported by: the Director, Office of Science, Office of High Energy Physics of the U.S. Department of Energy; the National Energy Research Scientific Computing Center, a DOE Office of Science User Facility; the U.S. National Science Foundation, Division of Astronomical Sciences; the National Astronomical Observatories of China, the Chinese Academy of Sciences and the Chinese National Natural Science Foundation. LBNL is managed by the Regents of the University of California under contract to the U.S. Department of Energy.\ The Photometric Redshifts for the Legacy Surveys (PRLS) catalog used in this paper was produced thanks to funding from the U.S. Department of Energy Office of Science, Office of High Energy Physics via grant DE-SC0007914.


\bibliography{journals,lit}

\onecolumngrid
\clearpage
\twocolumngrid

\emph{Fits and validations.}---Figure~\ref{fig:full_fit} shows the cross-power spectra measured between the Catalog 2 FRB DM and the DESI LIS North BGS sample over five photometric redshift ranges, with 10 equally spaced log bins over $40 < \ell < 8000$. This measurement was used to fit the theoretical model Eq.~(\ref{eq:cl_dg}) and to sample the parameter posterior distributions shown in Fig.~\ref{fig:posterior}. The fitting curves shown in Fig.~\ref{fig:full_fit} are computed from the median parameter values of their posterior distributions, with the $P_{eg}$ and $P_{fg}$ components shown separately. Using the Chi-square test to assess the goodness of fit
, we find that the $\chi^2$ per degree of freedom is 1.28 ($p$-value = 0.10), consistent with a good fit. Note that the cross-power spectra estimated by the catalog-based \texttt{NaMaster} estimator have a residual bandpower window function convolved. By computing this bandpower window function using \texttt{NaMaster} and forward modeling the signal template through it, we find the effect of the bandpower window on the signal is negligible given the $\ell$ bins in Fig.~\ref{fig:full_fit}.

The posterior distributions of $\langle \text{DM}_{\text{H}} \rangle$ in Fig.~\ref{fig:posterior} is highly skewed toward the upper bound of its prior, and a similar behavior can be seen with $\alpha$ and $z_*$. These priors are chosen on the basis of constraints from the CHIME/FRB Catalog 1 population study \citep{Shin_2023}. We investigate fitting these parameters with a more relaxed host DM prior $ 50 \text{ pc } \text{cm}^{-3} <\langle \text{DM}_{\text{H}} \rangle< 500 \text{ pc } \text{cm}^{-3}$ (as opposed to $ 50 \text{ pc } \text{cm}^{-3} <\langle \text{DM}_{\text{H}} \rangle< 250 \text{ pc } \text{cm}^{-3}$ in the original fit), and keep other priors the same. This results in the median values of $\langle \text{DM}_{\text{H}} \rangle$, $\alpha$, and $z_*$ being 385 $\text{pc } \text{cm}^{-3}$, -0.3, and 0.9, respectively, 
and their posterior distributions are no longer skewed. This large median $\langle \text{DM}_{\text{H}} \rangle$ value is only marginally preferred by the likelihood function, raising the detection significance from 5.1$\sigma$ to 5.2$\sigma$. 
We also find that the $k^{-1}_{cut}$ measurements are consistent regardless of whether we limit the host DM prior to within $250 \text{ pc } \text{cm}^{-3}$ or $500 \text{ pc } \text{cm}^{-3}$. In other words, the cross-correlation signal is insensitive to the host environment of FRBs and the background FRB population while robustly constraining the physical scale associated with feedback. 

\begin{figure*}
\includegraphics[width=2\columnwidth]{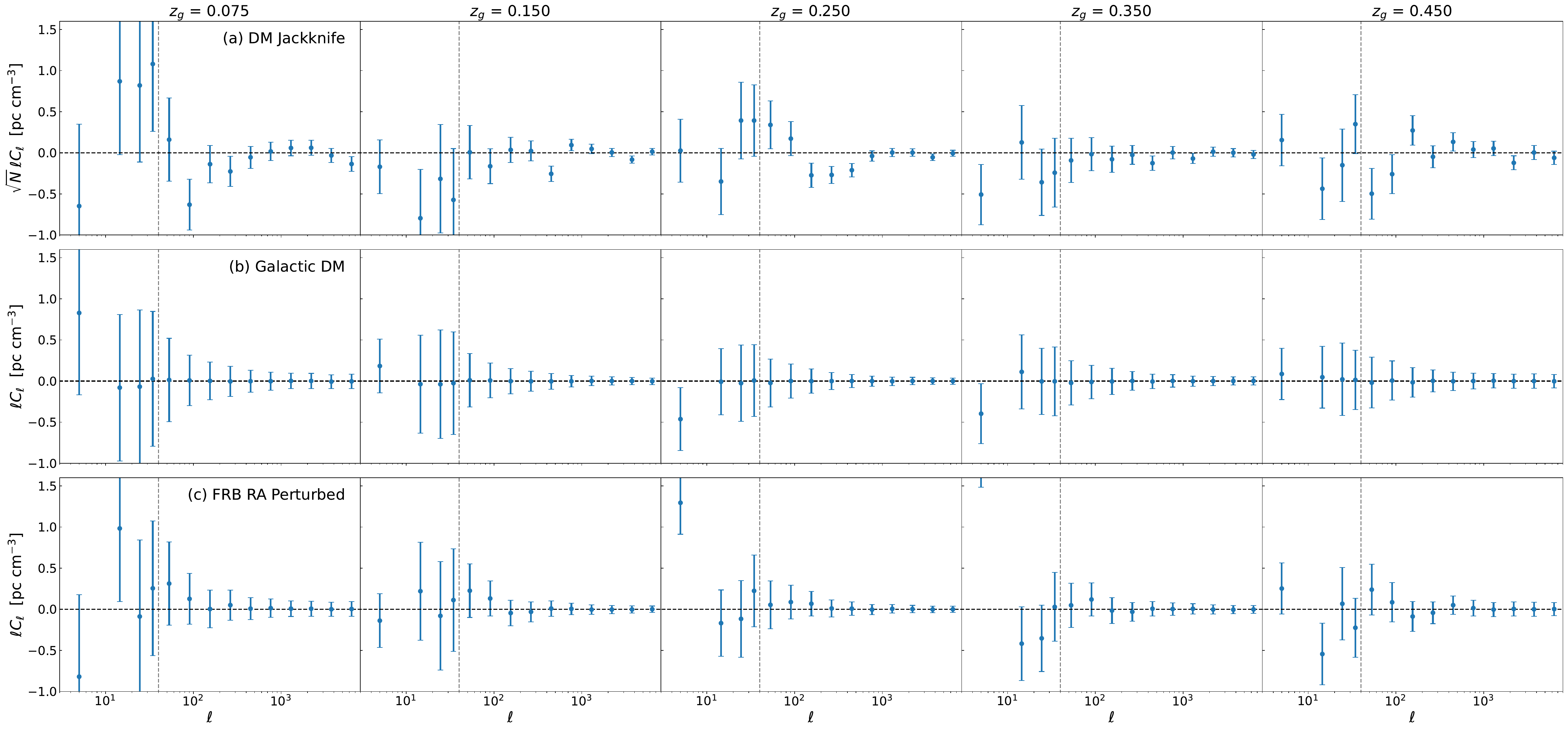}
\caption{\label{fig:null_tests}Null tests on the dispersion--galaxy cross-power measurement. (a) We randomly shuffle FRB DM while keeping their positions fixed and cross-correlate the shuffled DM with LIS BGS galaxies. The data shown are averaged over $N = 1000$ shuffles, and the error bars are errors on the average. Note that the amplitudes of the power spectra and their errorbars shown have been multiplied by a factor of $\sqrt{N}$. (b) We cross-correlate FRBs' Galactic DM predictions from NE2001 with LIS BGS galaxies. The error bars shown are those of the actual cross-power spectrum in Fig.~\ref{fig:full_fit}. (c) We randomly perturb FRB RA by $9^\circ$ and then cross-correlate their DM values with LIS BGS galaxies. The error bars shown are those of the actual cross-power spectrum. In all of the figures above, the measurements are consistent with zero detection at the angular scales used for fitting ($l > 40$, the regions to the right of the vertical dashed line).}
\end{figure*}

In addition, we have performed various consistency checks and null tests to ensure the robustness of our measurements.
Photometric surveys are susceptible to redshift errors and color-magnitude
depth variations, erasing power at small scales and introducing bias at large
scales. We therefore also measure the cross-power spectrum from DESI DR1
(Fig.~\ref{fig:full_fit}). Due to the limited coverage and number of samples,
the cross-power spectrum from DESI DR1 has a larger noise scatter but is
otherwise consistent with the measurement from DESI LIS\@. To check for systematic bias (spurious signal not from large-scale structures), we randomly shuffle the FRB DM values while keeping the FRB locations fixed. The randomly shuffled DMs no longer contain information on the intervening plasma and decorrelate with galaxies if no systematic bias is present. Our results averaged from 1000 random DM shuffles in Fig.~\ref{fig:null_tests}(a) are consistent with null detection.

Another potential source of systematics is inaccurate predictions for the Galactic DM contributions. We cross-correlate the Galactic DM predictions of FRBs from \texttt{NE2001} (subtracted off from the DM values in our real sample) with the galaxies and find no evidence of Galactic DM contributing to power at $\ell > 40$ (Fig.~\ref{fig:null_tests}(b)).   

Lastly, incomplete sky coverage of the FRB and galaxy surveys can cause power mixing between different angular scales, which can leak large-scale power to small scales. We therefore perturb the right ascension (RA) of FRBs by $9^\circ$ from their localized positions (we do not perturb the declinations (Dec) because CHIME has a sensitive Dec-dependent selection function). This erases small-scale power but maintains correlations at large scales. We observe that our measurement (Fig.~\ref{fig:null_tests}(c)) is consistent with zero at small scales ($\ell > 40$) after RA perturbation, indicating no power leakage from low $\ell$. This is consistent with studies \cite{Alonso_2019, Wolz_2025} that have shown that \texttt{NaMaster} power spectrum estimators can robustly compensate for power mixing due to survey geometries.

\onecolumngrid
\clearpage


\begin{center}
    \textbf{\large SUPPLEMENTARY MATERIAL}
\end{center}

\begin{center}
    \textbf{Dispersion--galaxy cross-power spectrum}
\end{center}

Here we present the full derivation of the dispersion--galaxy cross-power spectrum within the flat-sky and Limber approximation. Consider an FRB at position $\hat{n}\chi_f$, where $\hat{n}$ is the directional unit vector along the line of sight and $\chi_f$ is the comoving distance to the FRB\@. Assuming the Milky Way DM contribution can be subtracted and that FRBs on average have a uniform host DM contribution $\langle \text{DM}_\text{H} \rangle$, then by Eq.~(\ref{eq:dm_def}), the FRB DM can be modeled as
\begin{equation} \label{eq:single_dm}
    D(\hat{n}\chi_f) = \langle \text{DM}_\text{M} \rangle + \frac{\langle \text{DM}_\text{H} \rangle}{1 + z_f} + \int_0^{\chi_f} d\chi \frac{n_{e}(z)}{(1 + z)^2}  [1 + \delta_e (\hat{n}\chi, \chi)],
\end{equation}
where $\langle \text{DM}_\text{M} \rangle$ is the mean Milky Way halo DM contribution, $1 + z_f$ on the denominator of $\langle \text{DM}_\text{H} \rangle$ converts the host DM from the host galaxy's frame to the observer's frame (with $z_f$ being the redshift at $\chi_f$), $n_{e}(z)$ is the mean proper electron number density at a redshift $z$ that corresponds to the integration variable $\chi$, and $\delta_e (\hat{n}\chi, \chi)$ is the electron number overdensity at position $\hat{n}\chi$, with $\chi$ also serving as the time coordinate.

We now consider a catalog of such FRBs. For any line-of-sight direction $\hat{n}$, the averaged DM from all FRBs along that direction is
\begin{equation} \label{eq:dm_field}
    d(\hat{n}) = \frac{1}{N_f(\hat{n})} \int_0^\infty d\chi_f \chi_f^2 p_f(\chi_f) n_f(\hat{n}\chi_f, \chi_f)D(\hat{n}\chi_f),
\end{equation}
where $N_f(\hat{n})$ is the total number of detected FRBs along the $\hat{n}$ direction, $p_f(\chi_f)$ is the probability of detecting an FRB located at a comoving distance $\chi_f$ away, and $n_f(\hat{n}\chi_f, \chi_f)$ is the FRB number density. We want to express FRB number counts and number densities perturbatively, so we write 
\begin{equation} \label{eq:frb_3d_den}
    n_f(\hat{n}\chi_f, \chi_f) = n_{f0} [1 + \delta_f (\hat{n}\chi_f, \chi_f)],
\end{equation}
where $n_{f0}$ and $\delta_f (\hat{n}\chi_f, \chi_f)$ are the mean comoving FRB number density and FRB number overdensity, respectively. Similarly, we write
\begin{equation} \label{eq:frb_2d_den}
    N_f(\hat{n}) = N_{f0} [1 + \Delta f(\hat{n})],
\end{equation}
where $N_{f0}$ and $\Delta f(\hat{n})$ are the mean comoving FRB number count per line of sight and projected 2-dimensional FRB number overdensity, respectively, which can also be expressed as
\begin{align}
    N_{f0} = \int_0^\infty d \chi_f \chi_f^2 p_f(\chi_f) n_{f0},
\end{align}
and 
\begin{equation} \label{eq:Delta_f}
    \Delta f(\hat{n}) = \frac{1}{N_{f0}}\int_0^\infty d \chi_f \chi_f^2 p_f(\chi_f) n_{f0} \delta_f(\hat{n}\chi_f, \chi_f).
\end{equation}
Plugging Eqs.~(\ref{eq:single_dm}), (\ref{eq:frb_3d_den}), and (\ref{eq:frb_2d_den}) into Eq.~(\ref{eq:dm_field}), we can express the DM field in terms of perturbative quantities, namely $\delta_e (\hat{n}\chi, \chi)$, $\delta_f (\hat{n}\chi_f, \chi_f)$, and $ \Delta f(\hat{n})$, as
\begin{align} \label{eq:d_field_pert}
    d(\hat{n}) = \frac{1}{N_{f0}[1 + \Delta f(\hat{n})] } \int_0^\infty d\chi_f \chi_f^2 p_f(\chi_f) n_{f0} [1 + \delta_f (\hat{n}\chi_f, \chi_f)] \left\{ \langle \text{DM}_\text{M} \rangle + \frac{\langle \text{DM}_\text{H} \rangle}{1 + z_f} + \int_0^{\chi_f} d\chi \frac{n_{e}(z)}{(1 + z)^2} [1 + \delta_e (\hat{n}\chi, \chi)]\right\}.
\end{align}

We now define the \text{DM} overdensity as
\begin{align}
    \Delta d (\hat{n}) = d(\hat{n}) - \langle d(\hat{n}) \rangle, \label{eq:dm_overd_theory}
\end{align}
where the expectation $\langle\ldots\rangle$ denotes the ensemble average. Assuming the expectation of all overdensities in Eq.~(\ref{eq:d_field_pert}) goes to zero and keeping the overdensities to their lowest order, we obtain
\begin{equation}
    \Delta d(\hat{n}) = d_1(\hat{n}) + d_2(\hat{n}) + d_3(\hat{n}),
\end{equation}
where we have defined
\begin{equation} \label{eq:d1_real}
\begin{split}
    d_1(\hat{n}) & = -\frac{\Delta f(\hat{n})}{N_{f0}} \int_0^\infty d\chi_f \chi_f^2 p_f(\chi_f) n_{f0} \left[  \langle \text{DM}_\text{M} \rangle + \frac{\langle \text{DM}_\text{H} \rangle}{1 + z_f} + \int_0^{\chi_f} d\chi \frac{n_{e}(z)}{(1 + z)^2}  \right] \\
    & = -\Delta f(\hat{n}) \bar{d},
\end{split}
\end{equation}
where $\bar{d}$ is the average DM from all FRBs in the sample (and we assume that $\bar{d} = \langle d(\hat{n}) \rangle$ for a large sample of FRBs), 
\begin{equation} \label{eq:d2_real}
\begin{split}
    d_2(\hat{n}) & = \frac{1}{N_{f0}} \int_0^\infty d\chi_f \chi_f^2 p_f(\chi_f) n_{f0} \delta_f (\hat{n}\chi_f, \chi_f) \left[  \langle \text{DM}_\text{M} \rangle + \frac{\langle \text{DM}_\text{H} \rangle}{1 + z_f} + \int_0^{\chi_f} d\chi \frac{n_{e}(z)}{(1 + z)^2}  \right] \\
    & = \frac{1}{N_{f0}} \int_0^\infty d\chi_f \chi_f^2 p_f(\chi_f) n_{f0} \delta_f (\hat{n}\chi_f, \chi_f) \left[ \langle \text{DM}_\text{M} \rangle + \frac{\langle \text{DM}_\text{H} \rangle}{1 + z_f} +\langle \text{DM}_{\text{C}} (\chi_f) \rangle\right],
\end{split}
\end{equation}
where $\langle \text{DM}_{\text{C}} (\chi_f) \rangle$ is the Macquart relation, and finally
\begin{equation} \label{eq:d3_real}
    d_3(\hat{n})  = \frac{1}{N_{f0}} \int_0^\infty d\chi_f \chi_f^2 p_f(\chi_f) n_{f0} \int_0^{\chi_f} d\chi \frac{n_{e}(z)}{(1 + z)^2} \delta_e (\hat{n}\chi, \chi).
\end{equation}

Given a catalog of galaxies located within a thin shell centered at redshift $z_g$, we can count the number of galaxies along a line of sight $\hat{n}$ as
\begin{equation} \label{eq:g_field}
    g(\hat{n}) = \int d\chi \chi^2 p_g(\chi) n_{g0} [1 + \delta_g(\hat{n}\chi, \chi)],
\end{equation}
where $n_{g0}$ is the average comoving galaxy number density, $p_g(\chi)$ is the probability of detecting a galaxy at redshift $\chi$, and $\delta_g(\hat{n}\chi, \chi)$ is the galaxy number overdensity. We define the 2-dimensional projected galaxy number overdensity as
\begin{align}
    & \Delta g (\hat{n}) = \frac{g(\hat{n}) - \langle g(\hat{n})\rangle}{\langle g(\hat{n})\rangle}.
\end{align}
Using Eq.~(\ref{eq:g_field}), we find
\begin{equation} \label{eq:g_field_pert}
\begin{split}
    \Delta g(\hat{n}) & = \frac{1}{\langle g(\hat{n})\rangle} \int_0^\infty d\chi \chi^2 p_g(\chi) n_{g0} \delta_g(\hat{n}\chi, \chi) \\ & = \delta_g(\hat{n}\chi_g, \chi_g),
\end{split}
\end{equation}
where the second equality is reached by assuming that we detect all galaxies in a thin shell, namely $p_g(\chi) = 1$ around $\chi = \chi_g$ and 0 otherwise.

Having derived the real-space DM overdensity $\Delta d(\hat{n})$ and galaxy overdensity $\Delta g(\hat{n})$, we will next write down their Fourier transforms $\Delta d(\bm{l})$ and $\Delta g(\bm{l})$ within the flat-sky approximation, and then, using the Limber approximation, we will find the dispersion--galaxy cross-power spectrum through the Fourier space correlation function
\begin{equation} \label{eq:cl_dg_theory_def}
    \langle \Delta d(\bm{l}) \Delta g^*(\bm{l}^\prime) \rangle = (2\pi)^2 \delta^{(2)}(\bm{l} - \bm{l}^\prime) C^{dg}_l,
\end{equation}
where $\delta^{(2)}(\bm{l} - \bm{l}^\prime)$ is the two-dimensional Dirac delta function, $\bm{l}$ is the Fourier conjugate to the angular unit vector $\hat{n}$ in the flat-sky approximation, and $l = |\bm{l}|$.

We will first derive $\Delta g(\bm{l})$ since the expression for the galaxy number overdensity is simpler. Under the flat-sky approximation, we can write $\hat{n} = (\theta_x, \theta_y,1)$ with $\theta_x^2 + \theta_y^2 + 1^2 \approx 1$. Namely, the first two coordinates of $\hat{n}$ are along the sky plane with their coordinate values, $\theta_x$ and $\theta_y$, being small, and the last coordinate of $\hat{n}$ is perpendicular to the sky plane. Starting with the second line of Eq.~(\ref{eq:g_field_pert}), we express the real-space galaxy overdensity in terms of its Fourier transform as
\begin{equation}
\begin{split}
    \Delta g(\hat{n}) & = \int\frac{d^3k}{(2\pi)^3} e^{i\bm{k}\cdot\bm{x}} \delta_g(\bm{k}, \chi_g) \\
    & = \int\frac{d^3k}{(2\pi)^3} e^{i(k_x\theta_x + k_y\theta_y + k_z)\chi_g } \delta_g(\bm{k}, \chi_g) \\
    & = \frac{1}{(2\pi)^3} \int \frac{d l_x}{\chi_g} \frac{d l_y}{\chi_g} dk_z e^{i(l_x\theta_x + l_y\theta_y + k_z \chi_g) } \delta_g(\bm{k}, \chi_g) \\
    & = \int\frac{d^2 \bm{l}}{(2\pi)^2} e^{i \bm{l}\cdot\bm{\theta}} \frac{1}{\chi_g^2} \int \frac{dk_z}{2\pi} e^{i k_z \chi_g} \delta_g(\bm{k}, \chi_g),
\end{split}
\end{equation}
where on the first line, we have defined $\bm{x} = \hat{n}\chi_g = (\theta_x, \theta_y, 1)\chi_g$, and $\bm{k} = (k_x, k_y, k_z)$ is the Fourier dual to $\bm{x}$. To reach third line, we have defined $l_x \equiv k_x \chi_g$ and $l_y \equiv k_y \chi_g$, and on the final line, $\bm{l} \equiv (l_x, l_y)$ and $\bm{\theta} \equiv (\theta_x, \theta_y)$ are two-dimensional vectors. By the definition of the Fourier transform, we identify that
\begin{equation} \label{eq:g_field_fourier}
    \Delta g(\bm{l}) = \frac{1}{\chi_g^2}\int \frac{dk_z}{2\pi} e^{i k_z \chi_g} \delta_g(\bm{k}, \chi_g),
\end{equation}
where $\bm{k} = (l_x/\chi_g, l_y/\chi_g, k_z)$.

Following the same procedure, we can find the Fourier-space DM overdensities. From the second line of Eq.~(\ref{eq:d1_real}) and Eq.~(\ref{eq:Delta_f}), we find
\begin{equation} \label{eq:d1_fourier}
    d_1(\bm{l}) = -\frac{n_{f0}}{N_{f0}} \bar{d} \int_0^\infty d\chi_f \chi_f^2 p_f(\chi_f) \frac{1}{\chi_f^2}\int\frac{dk_z}{2\pi} e^{i k_z \chi_f} \delta_f(\bm{k}, \chi_f),
\end{equation}
where $\delta_f(\bm{k}, \chi_f)$ is the Fourier space FRB number overdensity. From Eq.~(\ref{eq:d2_real}), we find
\begin{equation} \label{eq:d2_fourier}
    d_2(\bm{l}) = \frac{1}{N_{f0}} \int_0^\infty d\chi_f \chi_f^2 p_f(\chi_f) n_{f0} \frac{1}{\chi_f^2} \int \frac{d k_z}{2\pi} e^{i k_z \chi_f} \delta_f(\bm{k}, \chi_f) \left[ \langle \text{DM}_\text{M} \rangle + \frac{\langle \text{DM}_\text{H} \rangle}{1 + z_f} +\langle \text{DM}_{\text{C}} (\chi_f) \rangle\right],
\end{equation}
and finally from Eq.~(\ref{eq:d3_real}), we write
\begin{equation} \label{eq:d3_fourier}
    d_3(\bm{l}) = \frac{1}{N_{f0}}  \int_0^\infty d\chi_f \chi_f^2 p_f(\chi_f) n_{f0}  \int_0^{\chi_f} d\chi \frac{n_{e}(z)}{(1 + z)^2} \int \frac{d k_z}{2 \pi} \frac{1}{\chi^2} e^{i k_z \chi} \delta_e (\bm{k}, \chi),
\end{equation}
where $\delta_e(\bm{k}, \chi)$ is the Fourier space electron number overdensity.

Next, we cross-correlate the DM overdensities with the galaxy number overdensity in the Fourier space as
\begin{equation} \label{eq:sum}
    \langle \Delta d(\bm{l}) \Delta g^*(\bm{l}^\prime) \rangle = \langle d_1(\bm{l}) \Delta g^*(\bm{l}^\prime) \rangle + \langle d_2(\bm{l}) \Delta g^*(\bm{l}^\prime) \rangle + \langle d_3(\bm{l}) \Delta g^*(\bm{l}^\prime) \rangle.
\end{equation}
We will compute each of the three terms on the right-hand side separately using the Limber approximation. From Eqs.~(\ref{eq:d1_fourier}) and (\ref{eq:g_field_fourier}), we write
\begin{equation} \label{eq:term_1}
    \langle d_1(\bm{l}) \Delta g^*(\bm{l}^\prime) \rangle = -\frac{n_{f0}}{N_{f0}} \bar{d} \int_0^\infty d\chi_f \chi_f^2 p_f(\chi_f) \frac{1}{\chi_f^2} \int \frac{dk_z}{2 \pi} e^{i k_z \chi_f} \int \frac{d k_z^\prime}{2\pi} \frac{1}{\chi_g^2} e^{-ik_z^\prime \chi_g} \langle \delta_f(\bm{k}, \chi_f) \delta_g^*(\bm{k}^\prime, \chi_g) \rangle. 
\end{equation}
Now we use the definition of the three-dimensional cross-power spectrum
\begin{equation} \label{eq:pfg_def}
    \langle \delta_f(\bm{k}, \chi_f) \delta_g^*(\bm{k}^\prime, \chi_g) \rangle = (2\pi)^3 \delta^{(3)}(\bm{k} - \bm{k}^\prime) P_{fg} (\bm{k}, \chi_f, \chi_g),
\end{equation}
where $\delta^{(3)}(\bm{k} - \bm{k}^\prime)$ is the three-dimensional Dirac delta function, and $P_{fg} (\bm{k}, \chi_f, \chi_g)$ is the unequal-time cross-power spectrum between FRB and galaxy number overdensities. 

Plugging Eq.~(\ref{eq:pfg_def}) into Eq.~(\ref{eq:term_1}) and recognizing that $
    \delta^{(3)}(\bm{k} - \bm{k}^\prime) = \delta^{(2)}\left( \frac{\bm{l}}{\chi_f} - \frac{\bm{l}^\prime}{\chi_g}\right) \delta(k_z - k_z^\prime)$, we can integrate Eq.~(\ref{eq:term_1}) over $k_z^\prime$ to obtain
\begin{equation} \label{eq:term_1_intermediate}
    \langle d_1(\bm{l}) \Delta g^*(\bm{l}^\prime) \rangle = -\frac{n_{f0}}{N_{f0}} \bar{d} \int_0^\infty d\chi_f \chi_f^2 p_f(\chi_f) \frac{2\pi}{\chi_f^2 \chi_g^2} \delta^{(2)}\left( \frac{\bm{l}}{\chi_f} - \frac{\bm{l}^\prime}{\chi_g}\right) \int dk_z e^{-i k_z (\chi_g - \chi_f) }P_{fg} (\bm{k}, \chi_f, \chi_g).
\end{equation}

Now we use the Limber approximation. Since the fields that we are correlating are two-dimensional projections, modes along the line-of-sight direction ($k_z$) are integrated over. Due to the oscillatory nature of Fourier modes in real space, modes with small wavenumber $\lambda_z \sim 1/k_z \ll \Delta \chi $, where $\Delta \chi$ is the range of integration, will be integrated to 0 along the $z$ direction. This means that $P_{fg} (\bm{k}, \chi_f, \chi_g)$ in the integrand will only contribute if $k_z \approx 0$, so we write
\begin{equation} \label{eq:limber}
    \int dk_z e^{-i k_z (\chi_g - \chi_f) }P_{fg} (\bm{k}, \chi_f, \chi_g) \approx P_{fg} (\bm{k}|k_z = 0, \chi_f, \chi_g) \int dk_z e^{-i k_z (\chi_g - \chi_f) } = 2\pi \delta(\chi_g - \chi_f) P_{fg} (\bm{k}|k_z = 0, \chi_f, \chi_g).
\end{equation}
With Eq.~(\ref{eq:limber}), we can integrate over $\chi_f$ in Eq.~(\ref{eq:term_1_intermediate}) to obtain
\begin{equation}
    \langle d_1(\bm{l}) \Delta g^*(\bm{l}^\prime) \rangle =  -\frac{n_{f0}}{N_{f0}} \bar{d} \, p_f(\chi_g) \frac{(2\pi)^2}{\chi_g^2} \delta^{(2)}\left( \frac{\bm{l}}{\chi_g} - \frac{\bm{l}^\prime}{\chi_g}\right)  P_{fg} (\bm{k}|k_z = 0, \chi_g).
\end{equation}
Now we use the Dirac delta function property $\delta^{(2)}\left( \frac{\bm{l}}{\chi_g} - \frac{\bm{l}^\prime}{\chi_g}\right) = \chi_g^2 \delta^{(2)}(\bm{l} - \bm{l}^\prime)$ to find
\begin{equation}
    \langle d_1(\bm{l}) \Delta g^*(\bm{l}^\prime) \rangle =  -(2\pi)^2 \delta^{(2)}(\bm{l} - \bm{l}^\prime) \frac{n_{f0}}{N_{f0}} p_f(\chi_g) P_{fg}(k = l/\chi_g, \chi_g) \bar{d},
\end{equation}
where we have assumed the power spectrum is isotropic, i.e., $ P_{fg}(\bm{k}, \chi_g) = P_{fg}(k, \chi_g) $.

With a change of variable on $p_f$,
\begin{equation} \label{eq:pf_zg}
    p_f(z_g) = \frac{1}{N_{f0}} \left. \frac{d N_f}{dz} \right \vert_{z_g} = \frac{1}{N_{f0}} \left. \frac{d N_f}{d\chi} \frac{d\chi}{dz}\right \vert_{z_g} = \frac{1}{N_{f0}} \chi_g^2 p_f(\chi_g) n_{f0} H^{-1}(z_g),
\end{equation}
where $H = (d\chi/dz)^{-1}$ is the Hubble parameter, we arrive at the final expression for the first term of the cross-correlation
\begin{equation} \label{eq:term_1_final}
    \langle d_1(\bm{l}) \Delta g^*(\bm{l}^\prime) \rangle = -(2\pi)^2 \delta^{(2)} (\bm{l} - \bm{l}^\prime) p_f(z_g) \frac{1}{\chi_g^2} H(z_g) P_{fg}(k = l/\chi_g, \chi_g) \bar{d}.
\end{equation}

Following the same argument, from Eqs.~(\ref{eq:d2_fourier}) and (\ref{eq:g_field_fourier}), we obtain 
\begin{equation}
\begin{split}
    \langle d_2(\bm{l}) \Delta g^*(\bm{l}^\prime) \rangle = \frac{n_{f0}}{N_{f0}} \int_0^\infty d\chi_f \chi_f^2 p_f(\chi_f) \frac{1}{\chi_f^2} \int \frac{d k_z}{2\pi} e^{i k_z \chi_f}  \left[ \langle \text{DM}_\text{M} \rangle + \frac{\langle \text{DM}_\text{H} \rangle}{1 + z_f} +\langle \text{DM}_{\text{C}} (\chi_f) \rangle\right]  \\ \times \int \frac{d k_z^\prime}{2\pi} \frac{1}{\chi_g^2} e^{-ik_z^\prime \chi_g} \langle \delta_f(\bm{k}, \chi_f) \delta_g^*(\bm{k}^\prime, \chi_g) \rangle.
\end{split}
\end{equation}
Now use Eq.~(\ref{eq:pfg_def}), integrate over $k_z^\prime$, and apply the Limber approximation. With the expression for $p_f(z_g)$ from Eq.~(\ref{eq:pf_zg}), we get
\begin{equation} \label{eq:term_2_final}
     \langle d_2(\bm{l}) \Delta g^*(\bm{l}^\prime) \rangle = (2\pi)^2 \delta^{(2)} (\bm{l} - \bm{l}^\prime) p_f(z_g) \frac{1}{\chi_g^2} H(z_g) P_{fg}(k = l/\chi_g, \chi_g) \left[ \langle \text{DM}_\text{M} \rangle + \frac{\langle \text{DM}_\text{H} \rangle}{1 + z_g} +\langle \text{DM}_{\text{C}} (\chi_g) \rangle\right].
\end{equation}

Finally, from Eqs.~(\ref{eq:d3_fourier}) and (\ref{eq:g_field_fourier}), we obtain
\begin{equation} \label{eq:d3_intermediate}
    \langle d_3(\bm{l}) \Delta g^*(\bm{l}^\prime) \rangle = \frac{n_{f0}}{N_{f0}}  \int_0^\infty d\chi_f \chi_f^2 p_f(\chi_f)  \int_0^{\chi_f} d\chi \frac{n_{e}(z)}{(1 + z)^2} \frac{1}{\chi^2}\int \frac{d k_z}{2 \pi}  e^{i k_z \chi} \int \frac{d k_z^\prime}{2\pi} \frac{1}{\chi_g^2} e^{-ik_z^\prime \chi_g} \langle \delta_e(\bm{k}, \chi) \delta_g^*(\bm{k}^\prime, \chi_g) \rangle,
\end{equation}
where
\begin{equation}
    \langle \delta_e(\bm{k}, \chi) \delta_g^*(\bm{k}^\prime, \chi_g) \rangle = (2\pi)^3 \delta^{(2)}\left( \frac{\bm{l}}{\chi} - \frac{\bm{l}^\prime}{\chi_g}\right) \delta(k_z - k_z^\prime) P_{eg} (\bm{k}, \chi, \chi_g),
\end{equation}
with $P_{eg} (\bm{k}, \chi, \chi_g)$ being the unequal time electron--galaxy cross-power spectrum.

Now, integrating Eq.~(\ref{eq:d3_intermediate}) over $k_z^\prime$ and applying the Limber approximation, we obtain
\begin{equation} \label{eq:term_3_final}
\begin{split}
    \langle d_3(\bm{l}) \Delta g^*(\bm{l}^\prime) \rangle
    & = (2\pi)^2 \delta^{(2)} (\bm{l} - \bm{l}^\prime) \frac{n_{e}(z_g)}{(1 + z_g)^2} \frac{1}{\chi_g^2} P_{eg} (k = l/\chi_g, \chi_g) \left[ \frac{1}{N_{f0}} \int_{\chi_g}^\infty d\chi_f \chi_f^2 p_f(\chi_f) n_{f0} \right] \\
    & = (2\pi)^2 \delta^{(2)} (\bm{l} - \bm{l}^\prime) \frac{n_{e}(z_g)}{(1 + z_g)^2} \frac{1}{\chi_g^2} P_{eg} (k = l/\chi_g, \chi_g) \frac{N_f(>\chi_g)}{N_f}.
\end{split}
\end{equation}
Note that on the first line of Eq.~(\ref{eq:term_3_final}), the lower bound of the integration over $\chi_f$ is $\chi_g$. This is due to the integration over $\chi$ after applying the Limber approximation is 0 if $\chi_f < \chi_g$. To reach the second line of Eq.~(\ref{eq:term_3_final}), we realize that
\begin{equation}
    N_f(>\chi_g) \equiv 4\pi\int_{\chi_g}^\infty d\chi_f \chi_f^2 p_f(\chi_f) n_{f0}
\end{equation}
is the total number of FRBs located beyond the galaxy redshift $z_g$, and $N_f = 4\pi N_{f0}$ is the total number of FRBs.

Putting Eqs.~(\ref{eq:term_1_final}), (\ref{eq:term_2_final}), and (\ref{eq:term_3_final}) into Eq.~(\ref{eq:sum}) and using the definition Eq.~(\ref{eq:cl_dg_theory_def}), we obtain the expression for the dispersion-galaxy cross-power spectrum as

\begin{equation}
        C_l^{dg} (z_g) = \frac{n_{e}(z_g)}{(1 + z_g)^2} \frac{f_f(z_g)}{\chi_g^2} P_{eg}(k = l/\chi_g, \chi_g) + \left[ \langle \text{DM}_\text{M} \rangle +\langle \text{DM}_{\text{C}} (\chi_g) \rangle  + \frac{\langle \text{DM}_\text{H} \rangle}{1 + z_g} - \bar{d}\right]
    \frac{p_f(z_g)}{\chi_g^2} H(z_g) P_{fg}(k = l/\chi_g, \chi_g),
\end{equation}
where $f_f(z_g) \equiv N_f(>\chi_g)/N_f$ is the fraction of FRBs located beyond the galaxy redshift $z_g$.

\clearpage

\begin{center}
    \textbf{Simple FRB population model based on the Schechter luminosity function}
\end{center}

We assume that the comoving number density of FRBs with an intrinsic luminosity $L$ follows the Schechter function \citep{schechter}:
\begin{equation} \label{eq:schecheter}
    dn(L) = \phi^*\left( \frac{L}{L_*} \right)^\alpha e^{-L/L_*} d\left( \frac{L}{L_*} \right),
\end{equation}
where $\phi^*$ is a normalization factor, and the parameters $L_*$ and $\alpha$, are the pivotal luminosity and power law index, respectively, to be determined from the data. To obtain the observed FRB number density at redshift $z$, we can integrate Eq.~(\ref{eq:schecheter}) from the minimum observable luminosity at redshift $z$, $L_{\text{min}}(z) = F_{\text{th}}4\pi d_L^2(z)$, where $F_{\text{th}}$ is the instrument sensitivity threshold flux and $d_L(z)$ is the luminosity distance to redshift $z$. We can further write $L_* = F_{\text{th}}4\pi d_L^2(z_*)$ and $L = F_{\text{th}}4\pi d_L^2(z^\prime)$, where $z_*$ and $z^\prime$ are parameterization parameters. Then Eq.~(\ref{eq:schecheter}) becomes
\begin{equation}
    dn(z^\prime) = \phi^* \left[ \frac{d_L^2(z^\prime)}{d_L^2(z_*)} \right]^\alpha e^{-d_L^2(z^\prime)/d_L^2(z_*)} \, d\left[ \frac{d_L^2(z^\prime)}{d_L^2(z_*)} \right].
\end{equation}
The observed FRB number density at redshift $z$ is then
\begin{equation}
    n(z) = \int^\infty_z dn(z^\prime),
\end{equation}
and the fraction of FRBs located beyond the galaxy redshift is
\begin{equation}
    f_f(z_g) = \frac{1}{N_{f,\text{total}}} \int^\infty_{\chi_g} 4 \pi \chi^2 n(\chi) d\chi,
\end{equation}
with the normalization condition $f_f(0)$ being 1, and $N_{f,\text{total}}$ is the total number of FRBs in the sample. Then the FRB redshift distribution is
\begin{equation}
    p_f(z_g) = -\left. \frac{d f_f}{dz} \right \vert_{z_g}.
\end{equation}

\begin{center}
    \textbf{Selecting the Bright Galaxy Sample from the North field of the DESI Legacy Imaging Survey}
\end{center}

\begin{SCfigure}[0.7][h]
\includegraphics[width=0.5\textwidth]{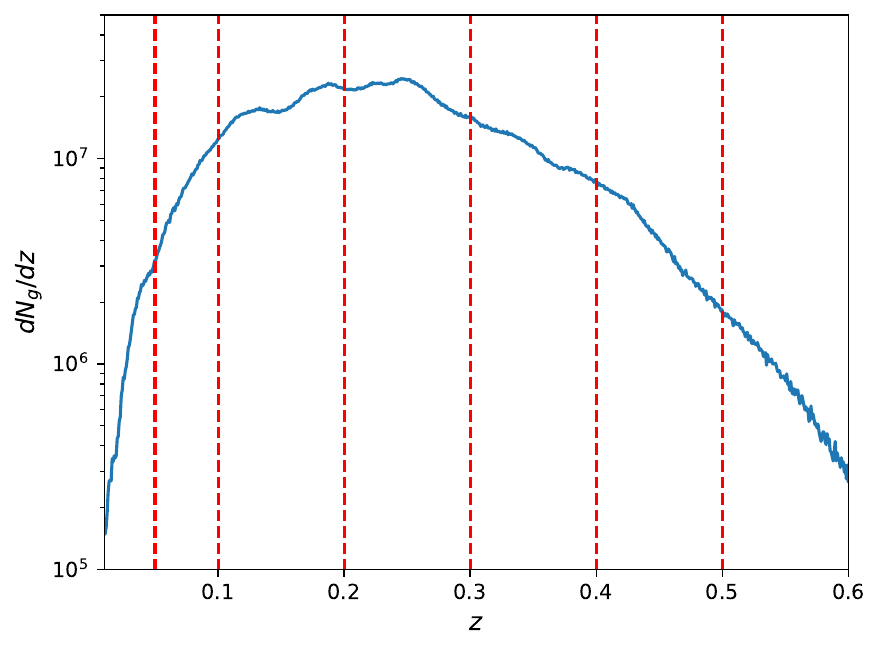}
\caption{\label{fig:gal_redshift} DESI LIS BGS North field galaxy photometric redshift distribution. The regions between the red dashed lines correspond to the five redshift bins of galaxies between $0.05 \leq z < 0.1$, $0.1 \leq z < 0.2$, $0.2 \leq z < 0.3$, $0.3 \leq z < 0.4$, and $0.4 \leq z < 0.5$, containing 374915, 1838042, 2116917, 1134574, and 434823 galaxies, respectively. }
\end{SCfigure}

The Dark Energy Spectroscopic Instrument (DESI) Legacy Imagining Survey (LIS) is a photometric galaxy survey designed to identify targets for the DESI spectroscopic survey. In this work, we select the Bright Galaxy Survey samples from LIS Data Release 8 (DR 8), based on the code infrastructure developed by \cite{Rafiei-Ravandi_2021} that followed the target selection criteria set out in \cite{Ruiz_Macias_2021}. In summary, the objects in the North field were observed jointly by the MzLS and BASS instruments, and their data were processed and stored in the ``sweep catalog'' as FITS files in the DESI LIS DR8 database \cite{LIS_DR8_north}. Following \cite{Ruiz_Macias_2021}, we apply an r-band magnitude cut ($r < 20$), mask sources contaminated by bright foreground stars, large galaxies, and globular clusters (corresponding to \texttt{MASKBIT} = 1, 5, 6, 7, 8, 9, 11, 12, 13, which is a stricter requirement than \cite{Ruiz_Macias_2021}), and only include sources that were observed at least twice (\texttt{NOBS} $\geq 2$) by each of the three color bands ($g$, $r$, and $z$) as required by \cite{zhou_2020}. We then remove objects identified as stars by DESI's source finding algorithm and apply the fiber magnitude cut and color cuts ($-1 < g-r<4$ and $-1 < r-z<4$, as defined by \cite{Ruiz_Macias_2021}) to identify the BGS samples. Lastly, we apply data quality cuts (\texttt{FRACMASKED\_i} $< 0.4$, \texttt{FRACIN\_i} $>0.3$, and \texttt{FRACFLUX\_i} $<5$ for the three bands $i = r,g,z$) to exclude targets with poor photometry. The resulting North-field BGS sample contains nearly 6 million galaxies. Using the photometric redshifts provided by \cite{zhou_2020}, we separate the BGS sample into five redshift bins (Fig.~\ref{fig:gal_redshift}) and cross-correlate the galaxies in each redshift bin with the DM values from the CHIME/FRB Catalog 2.

The DESI Collaboration provides random catalogs that characterize the survey geometry with no large-scale structure information. We apply the same target selection criteria to the random catalog objects, including \texttt{MASKBIT} cuts on foreground bright stars, large galaxies, and globular clusters, and cuts on the number of observations for each color band (requiring \texttt{NOBS} $\geq 2$). However, the random catalogs contain no information on color depth and redshift selection functions. In addition, the random catalogs provided by \cite{LIS_DR8_random} include sources from the entire DESI field. To isolate the sources in the North field, we collect the brick names (\texttt{BRICKNAME}) of all real North BGS sources, which is an identification number for each of the $0.25^\circ$ by $0.25^\circ$ sky patches observed by the survey. We keep the random sources with the same brick names as those of the real North field sources to assemble the random catalog for the BGS North sample. The resulting random catalog contains 40 times more sources than the real catalog. Due to the lack of redshift information, the same random catalog is used for all five photometric bins. Although the color depth and redshift selection functions are not characterized by the random catalog, we have shown in Fig.~\ref{fig:full_fit} that the cross-power spectrum measurements from the LIS are consistent with those from the DESI spectroscopic survey DR1, which has random catalogs that carefully characterize its selection functions.

\end{document}